\documentclass{entcs} 

\newcommand{\tr}[1]{}

\usepackage{entcsmacro}
\usepackage{graphicx}
\usepackage{amsmath}
\usepackage{amssymb}
\usepackage{stmaryrd}
\usepackage{mathpartir}
\usepackage{color}
\usepackage{MnSymbol}
\usepackage[figuresright]{rotating}
\usepackage{array}

\begin{document}

\def\lastname{Turon and Wand}
\begin{frontmatter}
  \title{A resource analysis of the $\pi$-calculus} 
  \author{Aaron Turon}
  \author{Mitchell Wand}
  \address{College of Computer and Information Science\\
Northeastern University\\
Boston MA, USA}
\begin{abstract}
We give a new treatment of the $\pi$-calculus based on the semantic
theory of separation logic, continuing a research program begun by
Hoare and O'Hearn.  Using a novel resource model that distinguishes
between public and private ownership, we refactor the operational
semantics so that sending, receiving, and allocating are commands that
influence owned resources.  These ideas lead naturally to two
denotational models: one for safety and one for liveness.  Both models
are fully abstract for the corresponding observables, but more
importantly both are very simple.  The close connections with the
model theory of separation logic (in particular, with Brookes's action
trace model) give rise to a logic of processes and resources.  
\end{abstract}
\end{frontmatter}

\newcommand{\CA}{\mathcal{A}}
\newcommand{\CO}{\mathcal{O}}
\newcommand{\CB}{\mathcal{B}}
\newcommand{\CM}{\mathcal{M}}
\newcommand{\CE}{\mathcal{E}}
\newcommand{\CS}{\mathcal{S}}
\newcommand{\CD}{\mathcal{D}}
\newcommand{\CJ}{\mathcal{J}}
\newcommand{\CL}{\mathcal{L}}
\newcommand{\CC}{\mathcal{C}}
\newcommand{\CP}{\mathcal{P}}
\newcommand{\CR}{\mathcal{R}}
\newcommand{\CT}{\mathcal{T}}
\newcommand{\CU}{\mathcal{U}}
\newcommand{\CN}{\mathcal{N}}
\newcommand{\CI}{\mathcal{I}}

\newcommand{\Sem}[2]{\left\llbracket #1 \right\rrbracket^{#2}}
\newcommand{\SemB}[1]{\left\llbracket #1 \right\rrbracket}

\newcommand{\LSem}[2]{\mathcal{L}\!\left\llbracket #1 \right\rrbracket^{#2}}
\newcommand{\LSemB}[1]{\mathcal{L}\!\left\llbracket #1 \right\rrbracket}

\newcommand{\ASem}[2]{\llparenthesis #1 \rrparenthesis #2}
\newcommand{\ASemB}[1]{\llparenthesis #1 \rrparenthesis }

\newcommand{\Ob}[2]{\mathcal{O}\!\left\llbracket #1 \right\rrbracket \! #2}
\newcommand{\ObB}[1]{\mathcal{O}\!\left\llbracket #1 \right\rrbracket }

\newcommand{\LOb}[2]{\mathcal{LO}\!\left\llbracket #1 \right\rrbracket \! #2}
\newcommand{\LObB}[1]{\mathcal{LO}\!\left\llbracket #1 \right\rrbracket }

\newcommand{\Int}[2]{\mathcal{I}\!\left\llbracket #1 \right\rrbracket \! #2}
\newcommand{\IntB}[1]{\mathcal{I}\!\left\llbracket #1 \right\rrbracket }

\newcommand{\eqdef}{\triangleq}
\newcommand{\ov}[1]{\ensuremath{\overline{#1}}}

\newcommand{\refines}{\sqsubseteq}
\newcommand{\abstracts}{\sqsupseteq}
\newcommand{\lub}{\sqcup}
\newcommand{\biglub}{\bigsqcup}
\newcommand{\glb}{\sqcap}
\newcommand{\bigglb}{\bigsqcap}

\newcommand{\new}{\textsf{new }}
\newcommand{\GA}{\ \ |\ \ }
\newcommand{\IF}{\textrm{if }}
\newcommand{\ra}{\rightarrow}
\newcommand{\Ra}{\Rightarrow}
\newcommand{\asm}{\blacktriangleright}

\newcommand{\pub}{\textsf{pub}}
\newcommand{\pri}{\textsf{pri}}
\newcommand{\known}{\textsf{known}}

\newcommand{\Pub}{\ \textsf{pub}}
\newcommand{\Pri}{\ \textsf{pri}}
\newcommand{\Known}{\ \textsf{known}}

\newcommand{\true}{\textsf{true}}
\newcommand{\false}{\textsf{false}}

\newcommand{\fault}{\lightning}

\newcommand{\ow}{\textrm{otherwise}}
\newcommand{\dom}{\textrm{dom}}
\newcommand{\fv}{\textrm{fv}}
\newcommand{\chans}{\textrm{chans}}
\newcommand{\newchans}{\textrm{newchans}}

\newcommand{\triple}[3]{\{#1\} #2 \{#3\}}

\newcommand{\cond}[3]{\textsf{if } #1 \textsf{ then } #2 \textsf{ else } #3}

\newcommand{\pref}{\triangleright}
\newcommand{\tparallel}{\parallel}

\newcommand{\step}[1]{\stackrel{#1}{\longrightarrow}}
\newcommand{\rstep}[1]{\stackrel{#1}{\rightarrowtriangle}}

\newcommand{\silent}{\textrm{silent}}

\newcommand{\pbegin}{\begin{tabular}[b]{l@{\quad}l}}
\newcommand{\pbegineqn}{\begin{tabular}[b]{rcll}}
\newcommand{\pend}{\end{tabular}}
\newcommand{\mc}[1]{\multicolumn{2}{l}{#1}}





\newcommand{\pcase}[1]{\vskip 5pt \noindent \mbox{\textit{Case: } \fbox{$#1$}}\quad}
\newcommand{\pcasec}[1]{\vskip 5pt \noindent \mbox{\textit{Case:} \fbox{\textsc{#1}} \ }}
\newcommand{\psubcase}[1]{\vskip 5pt \noindent \qquad \ \mbox{\textit{Subcase: } \fbox{$#1$}}\ }
\newcommand{\psubcasec}[1]{\vskip 5pt \noindent \qquad \mbox{\textit{Subcase:} \fbox{\textsc{#1}} \ }}

\newcommand{\psubsubcase}[1]{\vskip 5pt \noindent \qquad \quad \ \mbox{\textit{Subsubcase: } \fbox{$#1$}}\ }
\newcommand{\psubsubcasec}[1]{\vskip 5pt \noindent \qquad \quad \mbox{\textit{Subsubcase:} \fbox{\textsc{#1}} \ }}

\newcommand{\phave}[1]{$\begin{array}[t]{l} #1 \end{array}$}

\newcommand{\floor}[1]{\lfloor #1 \rfloor}

\newcommand{\proves}{\vdash}
\newcommand{\gives}{\proves}
\newcommand{\ok}{\checkmark}

\newcommand{\lref}[1]{lemma~\ref{lem-#1}}
\newcommand{\llref}[1]{\textrm{\small lem~\ref{lem-#1}}}
\newcommand{\sr}[1]{{\tiny\textsc{#1}}}

\newcommand{\rec}{\textsf{rec }}
\newcommand{\recB}{\textsf{rec}}

\newcommand{\dir}{\textrm{dir}}

\newcommand{\coind}{\mprset{fraction={===}}}
\newcommand{\ind}{\mprset{fraction={---}}}

\newcommand{\plist}[1]{\mbox{$\begin{array}[b]{l} #1 \end{array}$}}
\newcommand{\pllist}[1]{\mbox{$\begin{array}[b]{l@{\ :\ }l} #1 \end{array}$}}
\newcommand{\plistB}[1]{\mbox{$\begin{array}[c]{l} #1 \end{array}$}}

\definecolor{gray}{rgb}{0.9,0.9,0.9}
\newcommand{\gb}[1]{\colorbox{gray}{$#1$}}

\renewcommand{\RefTirName}[1]{\textsc{\tiny #1}}
\renewcommand{\TirName}[1]{\textsc{\tiny #1}}

\newcommand{\blocked}{\ \textrm{blocked}\ }

\newcommand{\tstep}[1]{\stackrel{#1}{\Longrightarrow}}
\newcommand{\tstepinf}[1]{\mathop{\Longrightarrow}^{#1}_\infty}

\newcommand{\nc}{\oplus}

\newcommand{\secref}[1]{\S\ref{sec:#1}}

\newcommand{\hbra}{
\hbox to .995 \columnwidth{\vrule width0.3mm height 1.8mm depth-0.3mm
                    \leaders\hrule height1.8mm depth-1.5mm\hfill
                    \vrule width0.3mm height 1.8mm depth-0.3mm}}
\newcommand{\hket}{
\hbox to .995 \columnwidth{\vrule width0.3mm height1.5mm
                    \leaders\hrule height0.3mm\hfill
                    \vrule width0.3mm height1.5mm}}

\newcommand{\hbraF}{
\hbox to .995 \textwidth{\vrule width0.3mm height 1.8mm depth-0.3mm
                    \leaders\hrule height1.8mm depth-1.5mm\hfill
                    \vrule width0.3mm height 1.8mm depth-0.3mm}}
\newcommand{\hketF}{
\hbox to .995 \textwidth{\vrule width0.3mm height1.5mm
                    \leaders\hrule height0.3mm\hfill
                    \vrule width0.3mm height1.5mm}}

\newcommand{\displayCap}[1]{\textbf{#1}}

\newenvironment{displayTab}[2][]{
  \noindent 
  \begin{minipage}{\columnwidth}
    \displayCap{#2}\hspace{\stretch{1}}\textit{#1}\\[-3ex]
}{
    \\[-1.5ex] 
  \end{minipage}
}

\newenvironment{display}[2][]{
  \vskip 4pt
  \noindent 
  \begin{minipage}{\columnwidth}
    \displayCap{#2}\hspace{\stretch{1}}\textit{#1}\\[-0.6ex]
    \hbra\\[-3ex]
}{
    \\[-3ex] \hket\\[-1.5ex]
  \end{minipage}
}

\newenvironment{eqdisplay}[2][]{
  \vskip 4pt
  \noindent 
  \begin{minipage}{\columnwidth}
    \displayCap{#2}\hspace{\stretch{1}}\textit{#1}\\[-0.6ex]
    \hbra\\[-3.5ex]
    \[ \begin{array}[c]{rcl}
}{
    \end{array} \]\\[-2ex] \hket\\[-1.2ex]
  \end{minipage}
}

\newenvironment{irdisplay}[2][]{
  \vskip 4pt
  \noindent  
  \begin{minipage}{\columnwidth}
    \displayCap{#2}\hspace{\stretch{1}}\textit{#1}\\[-0.6ex]
    \hbra\\[-2.5ex]
}{
    \\[-2ex] \hket\\[-1.2ex]
  \end{minipage}
}

\newcommand{\trip}[3]{\{#1\}\ #2\ \{#3\}}

\newcommand{\Pf}[1]{\textsf{Pf} [ #1 ] }

\newcommand{\leftlist}[1]{
  \left\{
  \begin{array}{l}
    #1
  \end{array}
  \right.
}

Names play a leading role in the $\pi$-calculus~\cite{Milner1992}:
they are both the means of communication, and the data communicated.
This paper presents a study of the $\pi$-calculus based on a new
mechanism for name management, which is in turn rooted in separation
logic.  The main benefit of this study is a very simple---but fully
abstract---denotational semantics for the $\pi$-calculus.

Traditionally, the use of names in the $\pi$-calculus is governed by
lexical, but dynamically-expandable, scope.  In the composite process
$P | \new x. Q$ for example, the channel $x$ is by virtue of scope
initially \emph{private} to $Q$.  The prefix $\new x$ is not an
imperative allocation.  It is a binder that remains fixed as $Q$
evolves---a constant reminder that $x$ is private---until $Q$ sends
$x$ in a message.  At that point, the binder is lifted to cover both
$P$ and $Q$, dynamically ``extruding'' the scope of $x$.  The
$\pi$-calculus relies on $\alpha$-renaming and side conditions about
freshness to ensure that its privacy narrative is borne out.



In contrast, work on separation logic has led to models of
dynamically-structured concurrency based on resources and ownership,
rather than names and scoping~\cite{Brookes2007,Calcagno2007}.  From
this perspective, programs consist of imperative commands that use
certain resources (their ``footprint'') while leaving any additional
resources unchanged.  Concurrent processes must divide resources
amongst themselves, with each process using only those resources it
owns.  Ownership makes it possible to constrain concurrent
interference, and thereby to reason compositionally about process
behavior.

In this paper, we reanalyze the $\pi$-calculus in terms of resources
and ownership, establishing a clear connection with models of
separation logic.  The analysis hinges on the use of resources to
specify not just that a process can do something, but that other
processes cannot.\footnote{ Such a reading of resources has already
  appeared in \emph{e.g.} deny-guarantee reasoning\cite{Dodds2009}.  }
Concretely, channels are resources that can be owned either publicly
or privately.  Public ownership asserts only that a channel can be
used by the owning process.  Private ownership asserts moreover that a
channel cannot be used by other processes.  And the prefix $\new x$
becomes an imperative action, allocating an initially private
channel.


Armed with this simple resource model, we give a new operational
semantics for the $\pi$-calculus~(\secref{operational}).  The
semantics is factored into two layers.  The first layer generates the
basic labeled transitions, without regard to their global
plausibility.  The second layer then uniformly interprets those labels
as resource transformers, filtering out implausible steps.  The
two-layer setup is reminiscent of Brookes's semantics for concurrent
separation logic~\cite{Brookes2007,Brookes2002}, and allows us to
blend message-passing and imperative interpretations of actions.


More importantly, the resource model also enables a very simple
denotational treatment of the $\pi$-calculus.  We give two
denotational interpretations, both trace-theoretic.  The
first~(\secref{safety}) captures safety properties only, while the
second~(\secref{liveness}) is also sensitive to divergence and some
branching behavior, along the lines of the failures/divergences model
with infinite traces~\cite{Roscoe1993}.  We prove that each model is
fully abstract with respect to  appropriate observables.

The semantic foundation reconciles the model theory of separation
logic with the $\pi$-calculus; what about the proof theory?  We sketch
an integration of separation logic with refinement calculus for
processes~(\secref{logic}).  Refinement is justified by the
denotational semantics, so the calculus is sound for contextual
approximation.  Resource reasoning allows us to derive an
\emph{interference-free expansion law} that uses privacy assertions to
rule out interference on a channel.


To provide an accurate model of the $\pi$-calculus, public/private
resources must be \emph{conservative} in a certain sense: once a
resource has been made public, it is impossible to make it private
again.  Work in separation logic has shown the usefulness of more
``aggressive'' resource models that capture not just what can and
cannot be done, but assert that certain things \emph{may} not be done.
We sketch a few such aggressive resource models~(\secref{resources}),
including an interpretation of fractional
permissions~\cite{Boyland2003} and of session types~\cite{Honda1998}.

Hoare and O'Hearn initiated a study of a $\pi$-calculus-like language
in terms of separation logic semantics~\cite{Hoare2008}.  That study
provided the impetus for our work, which goes farther by (1) handling
the full calculus, (2) handling liveness, (3) proving full abstraction
and (4) building a logic on the semantics.  There have also been
several fully abstract models of the
$\pi$-calculus~\cite{Stark2002,Hennessy2002,Fiore2002} based on
functor categories for modeling scope.  Our models
complement these by providing an elementary account of behavior,
structured around resources and abstract separation logic.  A full
discussion of related work is in~\secref{related}.


\section{A resource-driven operational semantics}
\label{sec:operational}

There are many variants of the $\pi$-calculus; here's ours:
\[
\begin{array}{c}
  P \ ::=\ \sum \pi_i. P_i \GA P \nc Q \GA \new x. P \GA P|Q \GA \rec X. P \GA X \\
  \pi \ ::=\ \ov{e}e' \GA e(x)  \qquad \qquad
  e \ ::=\ x \GA c 
\end{array}
\]
We distinguish between external choice ($+$) and internal choice
($\oplus$), which simplifies the liveness
semantics~(\secref{liveness}) but is not essential.  We also
distinguish between channels ($c, d$) and channel variables ($x, y, z$) and
include a simple grammar of channel expressions ($e$) ranging over
both.  A \emph{closed} process has no unbound channel or process
variables.  Closed processes may, however, refer to channel constants
and thereby communicate with an environment.

We write $0$ for an empty summation, which is an inert process.

\subsection{Generating actions}

The operational semantics of closed processes is given in two layers,
via two labelled transition systems.  In both systems, the
labels are (syntactic) \emph{actions}, given by the following grammar:
\begin{eqnarray*}
  \alpha &::=& c!d \GA c?d \GA \nu c \GA \tau \GA \fault \qquad (\textsc{Action})
\end{eqnarray*}
Actions record the concrete channels involved in sending, receiving,
and allocating, respectively.  The action $\tau$, as usual, represents
an internal (unobservable) step on the part of the process.  The
action $\fault$ represents a fault, caused by using an unowned
channel~(\secref{action-sem}).  Communication actions are dual: $\ov{c!d} = c?d$ and
$\ov{c?d} = c!d$, while $\ov{\nu c}$, $\ov{\tau}$, and $\ov{\fault}$
are undefined.

The first transition system generates all conceivable actions
associated with a process, without considering whether those actions
are globally plausible:

\begin{display}[$P \step{\alpha} Q$]{Operational semantics: action generation}
\[
\begin{array}{rcl}
\cdots + \ov{c}d.P + \cdots &\step{c!d}& P \tr{& \sr{ASend}}\\
\cdots + c(x).P +\cdots  &\step{c?d}& P \{d/x\} \tr{& \sr{ARecv}} \\
P_1 \nc P_2 &\step{\tau}& P_i \tr{& \sr{AChooseL/R}} \\ 
\new x.P &\step{\nu c}& P\{c/x\} \tr{& \sr{AAlloc}} \\
\rec X. P &\step{\tau}& 
  P\{\rec X.P/ X\}
  \tr{\multicolumn{2}{l}{P\{\rec X.P/ X\}}\ \sr{ARec}}
\end{array}
\quad
\begin{array}{c}
\infer[\tr{AIntL}]
  {P \step{\alpha} P'}
  {P|Q \step{\alpha} P'|Q}
\qquad
\infer[\tr{AIntR}]
  {Q \step{\alpha} Q'}
  {P|Q \step{\alpha} P|Q'}
\\
\\
\infer[\tr{ACom}]
  {P \step{\alpha} P' \\
   Q \step{\ov{\alpha}} Q'}
  {P|Q \step{\tau} P'|Q'}
\end{array}
\]
\end{display}
According to this semantics, we will have transitions like
\[
  \new x. \new y. \ov{x}y. 0
\ \step{\nu c}\
  \new y. \ov{c}y. 0
\ \step{\nu c}\
  \ov{c}c. 0
\ \step{c!c}\
  0
\]
where $c$ is allocated twice, and used to communicate with an
environment that cannot know it.  To filter out such executions, we
use resources.

\subsection{Resources and action semantics}\label{sec:action-sem}

The execution above is intuitively impossible because, after the first
$\nu c$ action, the process \emph{already owns} the channel $c$.
Similarly, for the process $\new x. \ov{x}x . 0$ the trace
\[
\new x. \ov{x}x. 0
\ \step{\nu c}\
  \ov{c}c. 0
\ \step{c!c}\
  0
\]
is impossible because the channel $c$, having just been allocated, is
unknown to the environment---so no parallel process could possibly be
on the other side of the communication, receiving along $c$.

Formally, resources are elements $\sigma$ of the domain $ \Sigma
\eqdef \textsc{Chan} \rightharpoonup \{ \pub, \pri \} $, where $\pub$
and $\pri$ are distinct atoms.  If a process is executing with
resources $\sigma$, it owns the channels $\dom(\sigma)$, and
$\sigma(c)$ tells, for each $c$, whether that ownership is exclusive.
Therefore, if $c \in \dom(\sigma)$, the action $\nu c$ is impossible.
Likewise, if $\sigma(c) = \pri$, the action $c!c$ is impossible.

The resources owned at a particular point in time determine not only
what is \emph{possible}, but also what is \emph{permissible}.  For
example, the process $\ov{c}d.0$ immediately attempts a communication
along the channel $c$.  If this channel is not allocated (\emph{i.e.},
not owned, \emph{i.e.}, not in $\dom(\sigma)$) then the process is
\emph{faulty}: it is attempting to use a dangling pointer.

We interpret actions $\alpha$ as \emph{resource transformers} of type
$\Sigma \ra \Sigma^\top_\bot$.\footnote{ The notation
  $\Sigma^\top_\bot$ denotes the set $\{\Sigma, \top, \bot\}$ and
  implies an ordering $\bot \leq \sigma \leq \top$ for all $\sigma \in
  \Sigma$.  The order structure follows abstract separation
  logic~\cite{Calcagno2007}, and is related to
  locality~(\secref{safety}).  }
Since all nondeterminism is resolved
during the generation of actions, these transformers are
deterministic.  A result of $\top$ or $\bot$ represents that an action
is not permissible or not possible, respectively.

Given the semantics $\ASem{\alpha} : \Sigma \ra \Sigma^\top_\bot$ of
actions (defined below), we can define a transition system that
\emph{executes} actions according to the currently-owned resources:
\begin{display}
[$P,\sigma \step{\alpha} P',\sigma'$]{Operational semantics: resource sensitivity}
\[
\infer[\tr{RStep}]
  {P \step{\alpha} P' \\
   \ASem{\alpha}{\sigma} =  \sigma'}
  {P,\sigma \rstep{\alpha} P', \sigma'}
\qquad
\infer[\tr{RFault}]
  {P \step{\alpha} P' \\
   \ASem{\alpha}{\sigma} = \top}
  {P,\sigma \rstep{\fault} 0,\sigma}
\]
\end{display}
Successful actions proceed normally, updating the owned
resources---note that if $\ASem{\alpha}{\sigma} = \sigma'$ then in
particular $\ASem{\alpha}{\sigma} \neq \top, \bot$.  Impermissible
actions noisily fail, generating the faulting label $\fault$.
Impossible actions silently fail to occur.





The semantics of actions is as follows:
\begin{display}[$\ASem{\alpha}{} : \Sigma \ra \Sigma^\top_\bot$]{Action semantics}
\[
\begin{array}[t]{rcl@{\quad}rcl}
  \ASem{c!d}{\sigma} &\eqdef&
    \begin{cases}
      \top & \{c,d\}\not\subseteq \dom(\sigma) \\
      \sigma[d \ \pub] & \sigma(c) = \pub \\
      \bot & \ow
    \end{cases} &
  \ASem{c?d}{\sigma} &\eqdef&
    \begin{cases}
      \top & c \notin \dom(\sigma) \\
      \sigma[d \ \pub] & \! 
        \begin{array}[t]{l}
          \sigma(c) = \pub,\\ \quad \sigma(d) \neq \pri 
        \end{array} \\
      \bot & \ow
    \end{cases} \\
  \ASem{\nu c}{\sigma} &\eqdef&
    \begin{cases}
      \sigma[c \ \pri] & c \notin\dom(\sigma) \\
      \bot & \ow
    \end{cases} &
  \ASem{\tau}{\sigma} &\eqdef& \sigma \qquad
  \ASem{\fault}{\sigma} \ \eqdef\ \top
\end{array}  
\]
\end{display}
Allocation is always permitted, but is not possible if the channel is
already allocated.  Allocated channels are initially private.  Sending
a channel publicizes it, but the communication is only possible if
performed over an already public channel, and only permitted over an
allocated channel.  A locally-unknown channel received from the
environment is known to the environment, and hence public; a
locally-known channel received from the environment cannot possibly
have been private.

\paragraph{Examples}

Consider the process $\new x. 0$.  We have
\[
  \new x. 0\quad \step{\nu c}\quad 0
\]
for every channel $c$.  It follows that
\[
  \new x. 0,\ \emptyset\quad \rstep{\nu c}\quad 0,\ [c \mapsto \pri]
\]
for every channel $c$, while executing with more resources
\[
  \new x. 0,\ [c \mapsto \pri]\quad \rstep{\nu d}\quad 0,\ [c \mapsto \pri] \uplus [d \mapsto \pri]
\]
results in constrained allocation: the $\uplus$ here denotes disjoint
union, meaning that $c \neq d$.  The fact that $c$ was already
allocated pruned one trace (preventing it from taking an impossible
step), but introduced no new traces.  Similarly,
\[
  \new x. \ov{x}x. 0\quad \step{\nu c} \ov{c}c.0\quad \step{c!c} 0
\]
but, taking resources into account, we have
\[
  \new x. \ov{x}x. 0,\ \emptyset\quad \rstep{\nu c}\quad
  \ov{c}c.0,\ [c \mapsto \pri]
\]
at which point the process is stuck: the action $c!c$ is prevented
from occurring, because $\ASem{c!c}[c\mapsto \pri] = \bot$.  This
deadlock is exactly what we expect to see when a process attempts to
communicate along a private channel.  Finally, we have
\[
  \new x. (\ov{x}x.0 | x(y).\ov{y}x.0)
\quad\step{\nu c}\quad
  \ov{c}c.0 | c(y).\ov{y}c.0
\quad\step{\tau}\quad
  0 | \ov{c}c.0
\quad\step{c!d}\quad
  0 | 0
\]
which, with resources, yields
\[
  \new x. (\ov{x}x.0 | x(y).\ov{y}x.0),\ \emptyset
\ \ \rstep{\nu c}\ \
  \ov{c}c.0 | c(y).\ov{y}c.0,\ [c \mapsto \pri]
\ \ \rstep{\tau}\ \
  0 | \ov{c}c.0,\ [c \mapsto \pri]
\]
Here we see that \emph{internal} communication along a private channel
is both possible and permitted: such internal steps appear as $\tau$
actions to the resource-sensitive stepping relation, and hence always
pass through.  On the other hand, the internal communication also
leaves the ownership of $c$ unchanged.  Because it remains private,
the final communication $\ov{c}c$ is stuck, as it should be.

\subsection{Process safety}

With the simple public/private resource model, faulting occurs only
when using an unallocated channel.  Our semantic framework can
accommodate deallocation, but doing so complicates the full
abstraction result, and we wish to focus on the standard
$\pi$-calculus.  Avoiding deallocation allows us to easily
characterize ``safe'' processes: we say $\sigma \proves P\ok$ iff $P$
is closed and all channel constants in $P$ are in $\dom(\sigma)$, and have:
\begin{lemma}
  If $\sigma \gives P \ok$ then
  $P,\sigma \stackrel{\fault}{\not\rightarrow}$, and
  if $P,\sigma \step{\alpha} P',\sigma'$ then $\sigma' \gives P' \ok$.
\end{lemma}






\section{Denotational semantics: safety traces}
\label{sec:safety}


Resources provide an intriguing refactoring of the operational
semantics for $\pi$-calculus, but their real payoff comes in the
elementary denotational model they support.  We begin with a simple
trace model capturing only (some) safety properties, which allows us
to focus on the role of resources.  Afterwards we incorporate
liveness~(\secref{liveness}) and its interaction with resources.

For the safety model, we have traces $t$, trace sets $T$ and behaviors
$B$:
\[
\begin{array}{c}
\textsc{Trace}
  \ \eqdef\ 
  \textsc{Action}^*
\qquad
\textsc{Beh}
  \ \eqdef\ 
  \Sigma \ra \textsc{TraceSet}
\\
\textsc{TraceSet} 
  \ \eqdef\ 
  \{ T\ :\ \emptyset \subset T \subseteq \textsc{Trace},\ T\ \textrm{prefix-closed} \}
\end{array}
\]
Processes will denote behaviors: sets of action traces determined by
the initially-available resources.  Not every action is observable.
We follow standard treatments of
$\pi$-calculus~\cite{Sangiorgi2001,Hennessy2002} in considering $\tau$
steps unobservable, and eliding $\nu c$ steps until just before the
allocated channel $c$ is sent over a public channel (a ``bound
send'').  Our denotational semantics shows that the operators of the
$\pi$-calculus are congruent for these observables, and the cited
works prove that similar observables are fully abstract for yet
coarser notions of observation.  The observables of an action $\alpha$
are a (possibly empty) trace, depending on the available resources:
\begin{display}
[$|\alpha|_\sigma : \textsc{Trace}$]
{Action observables}
\[
\begin{array}{rcl}
  |\tau|_\sigma &\eqdef& \epsilon \\
  |\nu c|_\sigma &\eqdef& \epsilon
\end{array} \qquad
\begin{array}{rcl}
  |\fault|_\sigma &\eqdef& \fault \\
  |c?d|_\sigma &\eqdef& c?d 
\end{array} \qquad
\begin{array}{rcl}
  |c!d|_\sigma &\eqdef& 
    \begin{cases}
      \nu d \cdot c!d & \sigma(d) = \pri \\
      c!d & \textrm{otherwise}
    \end{cases} 
\end{array}
\]
\end{display}
We write $t \cdot u$ or $tu$ for trace concatenation, and $\epsilon$
for the empty trace.  Although $\nu c$ is not immediately observable,
taking a $\nu c$ step affects the resources owned by the process, so
exposing $c$ later will cause the $\nu c$ step to reemerge.

The behavior of a process can be read from its operational semantics:
\begin{display}[$\ObB{P} : \textsc{Beh}$]{Safety observation}
\[
\infer
  { }
  {\epsilon \in \Ob{P}{\sigma}}
  \tr{\sr{OEps}}
\qquad
\infer
  {P,\sigma \rstep{\alpha} P', \sigma' \\
   t \in \Ob{P'}{\sigma'}}
  {|\alpha|_\sigma t \in \Ob{P}{\sigma}}
  \tr{\sr{OStep}}
\]
\end{display}
The goal of the denotational semantics is to calculate the same traces
compositionally over process structure.  

$\textsc{TraceSet}$ is a complete lattice under the subset order, and
behaviors inherit this order structure pointwise: we write $B \refines
B'$ if $B(\sigma) \subseteq B'(\sigma)$ for all $\sigma$ and have $(B
\lub B')(\sigma) = B(\sigma) \cup B'(\sigma)$.  The semantic
operators are monotonic (in fact, continuous), so we are justified
in defining \textsf{rec} as a fixpoint.  For the safety semantics,
which is based on finite observation, it is the least fixpoint.

The safety trace model is insensitive to branching behavior of
processes~\cite{Glabbeek1988}, so internal and external choice are
indistinguishable.  We interpret both forms of choice using $\lub$,
merging behaviors from all the alternatives.  For empty summations,
$\lub$ yields the smallest behavior: $\lambda \sigma. \{ \epsilon \}$.

The denotation function is parameterized by an environment $\rho$,
here taking channel variables $x$ to channels $c$, and process
variables $X$ to behaviors $B$.  It uses two additional operators,
$\pref$ and $\parallel$, which we will define shortly.
\begin{display}
[$\SemB{P} : \textsc{Env} \ra \textsc{Beh}$]
{Denotational semantics (safety)}
\[
\begin{array}{r@{\ \ \eqdef \ \ }r@{\ \pref\ }l}
  \Sem{\ov{e}e'.P}{\rho} & \rho  e ! \rho e' & \Sem{P}{\rho} \\
  \Sem{e(x).P}{\rho}    & \biglub_c \rho e ? c & \Sem{P}{\rho[x \mapsto c]} \\
  \Sem{\new x.P}{\rho}  & \biglub_c \nu c & \Sem{P}{\rho[x \mapsto c]} \\
  \Sem{\rec X.P}{\rho}   & \multicolumn{2}{r}{\mu B. \Sem{P}{\rho[X \mapsto B]}}
\end{array}
\qquad
\begin{array}{r@{\ \ \eqdef\ \ }l}
  \Sem{\sum \pi_i.P_i}{\rho} & \biglub_i \Sem{\pi_i.P_i}{\rho} \\
  \Sem{P \nc Q}{\rho}     & \Sem{P}{\rho} \lub \Sem{Q}{\rho} \\
  \Sem{P | Q}{\rho}     & \Sem{P}{\rho} \parallel \Sem{Q}{\rho} \\
  \Sem{X}{\rho}         & \rho(X) 
\end{array}
\]
\end{display}

The interpretation of prefixed processes resembles the operational
semantics: each clause of the denotational semantics generates all
locally-reasonable actions, without immediately checking global
plausibility.  We use $\lub$ to join the behaviors arising from each
action---once more reflecting nondeterminism---and we update the
environment as necessary.  The operator $\alpha \pref B$ prefixes an
action $\alpha$ to a behavior $B$ in a resource-sensitive way, playing
a role akin to the second layer of the operational semantics:
\begin{display}
[$\alpha \pref B : \textsc{Beh}$]
{Semantic prefixing}
\[
\begin{array}{r@{\ \ \eqdef\ \ }l}
(\alpha \pref B)(\sigma) &
         \{ \alpha t\ : \ASem{\alpha}{\sigma} = \sigma',\ t \in B(\sigma') \} 
\ \cup\  \{ \fault\ :\ \ASem{\alpha}{\sigma} = \top \} 
\ \cup\  \{ \epsilon \} 
\end{array}
\]
\end{display}
To maintain prefix-closure, we include $\epsilon$ as a possible trace.
A quick example: 
\[ 
  \Sem{\new x. \ov{x}x.0}{\emptyset}
\ =\
  \biglub_c \nu c \pref \Sem{\ov{x}x.0}{x \mapsto c}
\ =\
  \biglub_c \nu c \pref c!c \pref \Sem{0}{x \mapsto c}
\ =\
  \biglub_c \nu c \pref c!c \pref \lambda \sigma. \{ \epsilon \}
\]
This expansion of the definition resembles the traces we see from the
first layer of the operational semantics, without taking resources
into account.  The denotation, recall, is a \emph{behavior}: to
extract its set of traces, we must apply it to some particular
resource $\sigma$.  If we use the empty resource, we see that
\begin{eqnarray*}
  \left(\biglub_c \nu c \pref c!c \pref \lambda \sigma. \{ \epsilon \}\right)(\emptyset)
&=&
  \{ \epsilon \} \cup \bigcup_c \left\{ \nu c \cdot t\ :\ 
    t \in \left(c!c\pref \lambda \sigma. \{\epsilon\}\right)[c\mapsto \pri] \right\} \\
&=&
  \{ \epsilon \} \cup \bigcup_c \left\{ \nu c \cdot t\ :\ 
    t \in \{ \epsilon \} \right\} 
\end{eqnarray*}
in other words, we have $\Sem{\new x. \ov{x}x.0}{\emptyset}(\emptyset)
= \{ \epsilon \} \cup \bigcup_c \{ \nu c \}$.  Just as in the
operational semantics, the fact that $\ASem{c!c}[c \mapsto \pri] =
\bot$ prevents the $c!c$ step from being recorded.  Here, the prefix
closure (in particular, the inclusion of $\epsilon$ in every
application of $\pref$) ensures that we see the trace up to the point
that we attempt an impossible action.

Finally, we have parallel composition---the most interesting semantic
operator.  Here we must ask a crucial question for the denotational
semantics: if $\sigma$ is the resource belonging to $P|Q$, what
resources do we provide to $P$ and $Q$?  The question does not come up
in the operational semantics, which maintains a single, global
resource state, but a compositional semantics must answer it.

Consider the process $\new x.(\ov{x}c\ |\ x(z))$.  When the process
reaches the parallel composition, $x$ will still be private.  The
privacy of $x$ means that the subprocesses can only communicate with
each other (yielding $\tau$), not with the external environment of the
process.  But the subprocesses \emph{are} communicating with
environments external to themselves---namely, each other.  That is,
$x$ is private to $\ov{x}c\ |\ x(z)$, which cannot communicate along
it externally, but it is \emph{public} to the \emph{subprocesses}
$\ov{x}c$ and $x(z)$, which can.

Formally, we capture this narrative as follows: 
\begin{display}
[$B_1 \parallel B_2 : \textsc{Beh}$]
{Semantic parallel composition}
\[
\begin{array}{r@{\ \eqdef\ }l}
(B_1 \parallel B_2)(\sigma) &
  \bigcup_{t_i \in B_i(\widehat{\sigma})} (t_1 \tparallel t_2)(\sigma) 
\end{array}
\ \textrm{where}\ 
\widehat{\sigma}(c)\ \eqdef\
\begin{cases}
  \pub & c \in \dom(\sigma) \\
  \textrm{undefined} & \textrm{otherwise}
\end{cases}
\]
\end{display}
The resource $\sigma$ given to a parallel composition of behaviors is
fed in \emph{public-lifted} form ($\widehat{\sigma}$) to the composed
behaviors, yielding two sets of traces.  For each pair of traces $t_1$
and $t_2$ from these sets, we calculate all interleavings
$t_1 \mbox{$\tparallel$} t_2$:
\begin{display}
[$t \tparallel u : \textsc{Beh}$]
{Trace interleavings}
\[
\begin{array}{rcll}
t \tparallel u &\eqdef& 
  \lambda \sigma.\{\epsilon\}    &\IF t = \epsilon = u \\&\lub&
  \alpha \pref (t' \tparallel u) &\IF t = \alpha t' \\&\lub&
  \alpha \pref (t \tparallel u') &\IF u = \alpha u' \\&\lub&
  t' \tparallel u' &\IF t = \alpha t',\ u = \ov{\alpha} u'
\end{array}
\]
\end{display}
Interleaving at first glance appears standard, but note the use of
semantic prefixing $\pref$: \emph{the interleavings are not simply
  another set of traces, they are given as a \emph{behavior} that must
  be evaluated}.  We evaluate with the \emph{original} resources
$\sigma$.  The effect is that each interleaving is checked with
respect to the resources held by the \emph{combined} process.  This
additional check is the key to making the ``declare everything
public'' approach work, allowing us to take into account channels that
are private from the point of view of the combined process, but public
between the subprocesses.

An example helps illuminate the definitions: take the process
$\ov{d}c\ |\ d(z)$ with resources $\sigma = [c \mapsto \pub][d \mapsto
  \pri]$.  It is easy to
calculate that
\[
\begin{array}{l}
\Sem{\ov{d}c}{\emptyset}\!(\widehat{\sigma})\ =\ 
  \{ \epsilon, d!c \}
\qquad
\Sem{d(z)}{\emptyset}\!(\widehat{\sigma})\ =\ 
  \{ \epsilon \} \cup \{ d?e\ :\ e \in \textsc{Chan} \}
\\
d!c \parallel d?c 
\ =\ 
\left(d!c \pref d?c \pref \lambda \sigma. \{\epsilon\}\right)
\lub
\left(d?c \pref d!c \pref \lambda \sigma. \{\epsilon\}\right)
\lub
\left(\lambda \sigma. \{\epsilon\}\right)
\end{array}
\]
The interleaving $d!c \parallel d?c$ includes the case that $d!c$ and
$d?c$ are two sides of the same communication (yielding $\lambda
\sigma. \{\epsilon\}$) and the two possible orderings if they are not.
From the point of view of $\widehat{\sigma}$, which has lost the
information that $d$ is private to the combined process, this is the
most we can say.  However, the interleaving is built using the
prefixing operation $\pref$, so when we evaluate it with respect to
the original $\sigma$, some traces will be silently dropped:
\begin{eqnarray*}
&&      (d!c \parallel d?c)(\sigma)\\
&\ =\ &        
        (d!c \pref d?c \pref \lambda \sigma. \{\epsilon\})(\sigma)
        \cup 
        (d?c \pref d!c \pref \lambda \sigma. \{\epsilon\})(\sigma)
        \cup
        (\lambda \sigma. \{\epsilon\})
(\sigma) \\
&\ =\ &        
        \{ \epsilon \} \cup
        \{ \epsilon \} \cup
        \{ \epsilon \}
\end{eqnarray*}
In particular, for any $B$ we have 
$
(d!c \pref B)(\sigma) = (d?c \pref B)(\sigma) = \{ \epsilon \}
$
because $\sigma(d) = \pri$.  We are left only with traces that could
arise from internal communication, as expected.  That is, 
$\Sem{\new x.(\ov{x}c|x(y))}{\emptyset}[c \mapsto\pub] = \{ \epsilon \}$.
More generally, we can show 
$\Sem{\new x.(\ov{x}c|x(y))}{\emptyset}\sigma = \Sem{0}{\emptyset}\sigma$
whenever $c \in \dom(\sigma)$.

Because $\ASem{\fault}{\sigma} = \top$, we have $\fault \pref B =
\lambda \sigma. \{ \fault, \epsilon \}$ for any $B$.  Thus, when a
$\fault$ action is interleaved, the interleaving is terminated with
that action.

In summary, we calculate the traces of $P|Q$ by calculating the traces
of $P$ and $Q$ under conservatively public-lifted resources, then
evaluating the interleavings with complete information about what
resources $P|Q$ actually owns.



\paragraph{Example calculations}

Before proving full abstraction, we briefly examine a few of the
expected laws.  For example, why does $\SemB{\new x.0} = \SemB{0}$?
Expanding the former, we get $\biglub_c \nu c \pref \lambda \sigma
. \{ \epsilon \}$.  When applied to a particular $\sigma$, this
behavior yields the simple set $\{ \epsilon \}$, because $|\nu
c|_\sigma = \epsilon$.  This simple example sheds light on the
importance of action observation $|-|$: it is crucial for ignoring
when, or in some cases whether, channels are allocated.

A more complex example is the following:
\begin{eqnarray*}
  \Sem{\new x.\new y.P}{\rho}
&=&
  \biglub_c \nu c \pref \Sem{\new y.P}{\rho[x \mapsto c]} \\
&=&
  \biglub_c \nu c \pref \biglub_d \nu d \pref \Sem{P}{\rho[x \mapsto c, y \mapsto d]} \\
&=&
  \biglub_{c,d} \nu c \pref \nu d \pref \Sem{P}{\rho[x \mapsto c, y \mapsto d]} \\
&=&
  \biglub_{c,d} \nu d \pref \nu c \pref \Sem{P}{\rho[x \mapsto c, y \mapsto d]} \\
&=&
  \biglub_d \nu d \pref \biglub_c \nu c \pref \Sem{P}{\rho[x \mapsto c, y \mapsto d]} \\
&=&
  \biglub_d \nu d \pref \Sem{\new x.P}{\rho[y \mapsto d]}
\ =\
  \Sem{\new y.\new x.P}{\rho}
\end{eqnarray*}
The key step is swapping $\nu c$ and $\nu d$, which relies on the
lemma $\nu c \pref \nu d \pref B = \nu d \pref \nu c \pref B$.  The
validity of this lemma, again, relies on observability: $|\nu c|_\sigma
= |\nu d|_\sigma = \epsilon$ for all $\sigma$.

\subsection{Congruence for the basic operators}

We prove full abstraction by proving a \emph{congruence} result for
each operator in the language.  For the operators other than parallel
composition, we show:

\begin{lemma}[Core congruences]
All of the following equivalences on closed processes hold:
\begin{enumerate}
\item $\ObB{0} = \Sem{0}{\emptyset}$
\item $\ObB{\ov{c}d.P} = c!d \pref \ObB{P}$
\item $\ObB{c(x).P} = \biglub_d c?d \pref \ObB{P\{d/x\}}$
\item $\ObB{\new x.P} = \biglub_c \nu c \pref \ObB{P\{c/x\}}$
\item $\ObB{\sum_i P_i} = \biglub_i \ObB{P_i}$
\item $\ObB{P \nc Q} = \Ob{P} \lub \ObB{Q}$
\end{enumerate}
\end{lemma}
\noindent
These equivalences are straightforward to show; we prove each by
showing containment in both directions.  For illustration, we give the
proof that $\ObB{c(x).P} \subseteq \biglub_d c?d \pref \ObB{P\{d/x\}}$:
\begin{proof}
Let $\sigma \in \Sigma$ and $t \in \Ob{c(x).P}{\sigma}$.  
We analyze cases on the derivation of $t \in \Ob{c(x).P}{\sigma}$:

\pcase{
  \infer
    {\phantom{a} }
    {\epsilon \in \Ob{c(x).P}{\sigma}}
}

\vskip 2pt\noindent
Let $d$ be a channel.  Then $t = \epsilon \in c?d \pref
\ObB{P\{d/x\}}$ by definition of $\pref$.  The result follows by
monotonicity of $\lub$.

\pcase{
  \infer
    {c(x).P,\sigma \step{\alpha} P',\sigma' \\ t' \in \Ob{P'}{\sigma'}}
    {|\alpha|_\sigma t' \in \Ob{c(x).P}{\sigma}}
}

\vskip 2pt

Reasoning by inversion, we see that there are two subcases:

\psubcase{\exists d.\ \alpha = c?d,\ \ASem{c?d}{\sigma} = \sigma',\ P' = P\{d/x\}}

\vskip 4pt

Then $t = \alpha t' \in \biglub_d c?d \pref \Ob{P\{d/x\}}$ trivially
by the definition of $\pref$.

\psubcase{\alpha = \fault,\ c \notin\dom(\sigma),\ P' = 0}

\vskip 4pt

Then $t = \alpha t' = \fault$ because $\Ob{0}{\sigma'} = \{ \epsilon
\}$.  That $\fault \in \biglub_d c?d \pref \Ob{P\{d/x\}}$ again
follows easily by the definition of $\pref$.
\end{proof}

\subsection{Congruence for parallel composition}

The justification of our treatment of parallel composition goes back
to the intuitions from the beginning of the paper: concurrent process
must divide resources amongst themselves, with each process using only
those resources it owns.  We say $\sigma$ separates into $\sigma_1$
and $\sigma_2$ if the following conditions hold:
\begin{display}[$(\sigma_1 \parallel \sigma_2) \subseteq \Sigma$]{Parallel separation}
\[
  \sigma \in (\sigma_1 \parallel \sigma_2) \ \eqdef\
  \left\{
  \begin{array}{l}
    \dom(\sigma) = \dom(\sigma_1) \cup \dom(\sigma_2) \\
    \sigma_1(c) = \pri \implies \sigma(c)=\pri,\ c\notin\dom(\sigma_2) \\
    \sigma_2(c) = \pri \implies \sigma(c)=\pri,\ c\notin\dom(\sigma_1) 
  \end{array}
  \right.
\]
\end{display}

We understand this definition as saying: if $\sigma_1$ and $\sigma_2$
are resources separately held by $P$ and $Q$ respectively, then
$\sigma$ is \emph{possibly} the resource held by $P|Q$.  The
subresources $\sigma_i$ do not uniquely determine a combination
$\sigma$ because resources public to the subprocess may, or may not,
be private to the combined process.\footnote{
  This means that $\Sigma$ with $\parallel$ does not form a separation 
  algebra~\cite{Calcagno2007}; see~\secref{resources}.
}
Separation crisply captures the desired meaning of public and private
ownership: if one subprocess owns a resource privately ($\sigma_1(c) =
\pri$), then the other subprocess does not own the resource at all ($c
\notin\dom(\sigma_2)$), but both processes may own a resource
publicly.
%


To show that that $\ObB{P_1|P_2} = \ObB{P_1} \mbox{$\parallel$}
\ObB{P_2}$, we must show that our strategy of interleaving traces from
publicly-lifted resources agrees with the global operational
semantics.  A key idea is that $\sigma \in \sigma_1 \parallel
\sigma_2$ constitutes an invariant relationship between the resources
owned by subprocesses (in the denotational semantics) and those owned
by the composite process (in the operational semantics).  The
invariant holds initially because $\sigma \in \widehat{\sigma}
\parallel \widehat{\sigma}$.


The unobservability of $\nu c$ steps complicates matters somewhat: it
means there is an additional perspective on resources---call it
$\sigma_{\rm den}$---owned by a composite process.  Generally,
$\sigma_{\rm den}$ underestimates the true resources $\sigma$ of the
operational semantics.  Consider the denotational interleaving of two
traces $t_1$ and $t_2$ from subprocesses $P_1$ and $P_2$ respectively.
If $P_1$ allocates a channel, that allocation does not appear
immediately in $t_1$, and hence does not appear immediately in the
resources $\sigma_{\rm den}$ of the interleaving, while it would
appear in $\sigma$ operationally.  During denotational interleaving,
the same channel can even be owned privately in \emph{both} $\sigma_1$
and $\sigma_2$.  The key observation here is that either both
subprocesses eventually reveal a given private channel---in which case
the denotational interleaving is filtered out---or at least one
subprocess does not---in which case its choice of channel is
irrelevant.  Altogether, the four resources---$\sigma_{\textrm{op}}$,
$\sigma_{\textrm{den}}$, $\sigma_1$, and $\sigma_2$---are always
related: 
\[
\CI(\sigma_{\textrm{op}}, \sigma_{\textrm{den}}, \sigma_1, \sigma_2)\ \eqdef\ 
\sigma_{\textrm{op}} \in \sigma_1\parallel\sigma_2,\ 
\sigma_{\rm den} = \sigma_{\textrm{op}} \setminus \{ c\ :\ \sigma_1(c) = \pri \vee \sigma_2(c) = \pri \}
\]






Validating parallel composition requires another important lemma,
\emph{locality} from abstract separation
logic~\cite{Calcagno2007}.\footnote{ For simplicity we avoid the
  order-theoretic definition here, which requires lifting some of our
  constructions to $2^\Sigma$ in a way that is not otherwise useful.
}

\begin{lemma}[Locality]
  If $\sigma \in \sigma_1 \parallel \sigma_2$ then
\begin{itemize}
  \item if $\ASem{\alpha}{\sigma} = \top$ then $\ASem{\alpha}{\sigma_1} = \top$, and
  \item if $\ASem{\alpha}{\sigma} = \sigma'$ then
    $\ASem{\alpha}{\sigma_1} = \top$ or $\ASem{\alpha}{\sigma_1} =
    \sigma'_1$ for some $\sigma'_1$ with $\sigma' \in \sigma'_1
    \parallel \sigma_2$.
\end{itemize}
\end{lemma}
The lemma characterizes the transformations an action can make given
some composite resources $\sigma$ in terms of its behavior on
subresources $\sigma_1$.  Providing additional resources can never
introduce new faults, and if the action does not fault given just
$\sigma_1$ resources, then the changes it makes to $\sigma$ must only
change the $\sigma_1$ portion (framing).

Locality was introduced to characterize the frame rule of separation
logic~\cite{Calcagno2007}, but we use it here to characterize
interleaving steps in parallel composition.  We have a related lemma
for internal communication steps:

\begin{lemma}[Communication]
  If $\sigma \in \sigma_1 \parallel \sigma_2$,
$\ASem{\alpha}{\sigma_1} = \sigma'_1$ and $\ASem{\ov{\alpha}}{\sigma_2} = \sigma'_2$ then 
    $\sigma \in \sigma'_1 \parallel \sigma'_2$.
\end{lemma}

We prove each direction of congruence separately:
\begin{lemma}
  If $\CI(\sigma_{\textrm{op}}, \sigma_{\textrm{den}}, \sigma_1, \sigma_2)$,
  $\sigma_i \proves P_i \ok$ and $t \in \Ob{P_1|P_2}{\sigma_{\textrm{op}}}$
  then\\ $t \in (t_1\parallel t_2)(\sigma_{\textrm{den}})$ for some
  $t_i \in \Ob{P_i}{\sigma_i}$.
\end{lemma}
\begin{lemma}
  If $\CI(\sigma_{\textrm{op}}, \sigma_{\textrm{den}}, \sigma_1, \sigma_2)$,
  $\sigma_i \proves P_i \ok$, $t_i \in \Ob{P_i}{\sigma_i}$, and \\
  $t \in (t_1\parallel t_2)(\sigma_{\textrm{den}})$ then 
$t \in \Ob{P_1|P_2}{\sigma_{\textrm{op}}}$.
\end{lemma}
The first of these two lemmas is easier to prove, because we are given
a trace $t$ derived from the operational semantics of the composite
processes.  This means that the subprocesses are guaranteed not to
independently allocate the same channel.  The second lemma requires
more care, using the insights mentioned above about renaming unexposed
channels.

The assumptions $\sigma_i \proves P_i \ok$ are needed to ensure that
the processes we are working with do not fault.  The reason that
faulting is problematic is seen in the following example:
\begin{eqnarray*}
&&  \new x.\ov{c}x.0\ |\ c(y).\ov{c}y.\ov{d}y.0),\ [c \mapsto \pub] \\
&\rstep{\nu d}\ \ &
  \ov{c}d.0\ |\ c(y).\ov{c}y.\ov{d}y.0,\ [c \mapsto \pub, d \mapsto \pri] \\
&\rstep{\tau}\ \ &
  0\ |\ \ov{c}d.\ov{d}c.0,\ [c \mapsto \pub, d\mapsto\pri] \\
&\rstep{c!d}\ \ &
  0\ |\ \ov{d}c.0,\ [c \mapsto \pub, d\mapsto\pub] \\
&\rstep{d!c}\ \ &
  0\ |\ 0,\ [c \mapsto \pub, d\mapsto\pub] 
\end{eqnarray*}
The uncomfortable aspect of this derivation is that the channel $d$
occurred in the process initially, even though it was not owned.  As a
result, the process was able to \emph{allocate} $d$, in a sense
falsely capturing the constant $d$ that initially appeared.  In cases
where the process allocates a different channel than $d$, it will
fault when it attempts to communicate along the constant channel $d$.
But in this ``lucky'' case, the operational semantics allows
communication along the constant channel.

The denotational semantics, however, \emph{always} generates a fault.
It computes the traces compositionally, meaning that a channel $d$
allocated by one subprocess is not immediately available for use by a
parallel subprocess.

Our full abstraction result applies only to nonfaulty processes,
which, fortunately, is a trivial syntactic check.  However, this does
limit its applicability to languages that include features like
deallocation, which makes checking for safety more difficult.



\subsection{Full abstraction}

To complete the proof of full abstraction, we must deal with
recursion.  We begin with the usual unwinding lemma, proved in the
standard syntactic way:

\begin{lemma}[Unwinding]
  We have
  $\ObB{\rec X.P} = \biglub_n \Ob{\recB_n X.P}$, where $\recB_0 X.P \eqdef
  \rec X. X$ and $\recB_{n+1} X.P \eqdef P\{\recB_n X.P/X\}$.
\end{lemma}

We also have the standard substitution lemmas:
\begin{lemma}[Substitution]
  We have
  $\Sem{P[Q/X]}{\rho} = \Sem{P}{\rho[X \mapsto Q]}$ and\\
  $\Sem{P[c/x]}{\rho} = \Sem{P}{\rho[x \mapsto c]}$.
\end{lemma}

\noindent
Combined these lemmas with the previous congruence results, it is
straightforward to show the following theorem relating the observed
operational traces to those calculated denotationally:

\begin{theorem}[Congruence]
  If $P$ closed and $\sigma \gives P \ok$ then
  $\Ob{P}{\sigma} = \Sem{P}{\emptyset}\sigma$.
\end{theorem}
\noindent
To prove this theorem, we must generalize it to deal with open terms.
We do this by introducing a \emph{syntactic environment} $\eta$ as a
finite map taking channel variables to channels and process variables
to closed processes.  Given a syntactic environment $\eta$ the
corresponding semantic environment $\widehat{\eta}$ is given by:
\[
  (\widehat{\eta})(x)\ \eqdef\ \eta(x) \qquad
  (\widehat{\eta})(X)\ \eqdef\ \ObB{\eta(X)}
\]
We write $\eta P$ for the application of $\eta$ as a syntactic
substitution on $P$.  The needed induction hypothesis for congruence
is then 
\begin{center}
if $\sigma \gives \eta P \ok$ then $\Ob{\eta P}{\sigma} = \Sem{P}{\widehat{\eta}}\sigma$.
\end{center}

Define $P =_{\textsc{Den}} Q$ iff $\Sem{P}{\rho}\sigma =
\Sem{Q}{\rho}\sigma$ for all $\sigma$ such that $\sigma \gives P\ok$
and $\sigma \gives Q\ok$.  Likewise, let $P =_{\textsc{Op}} Q$ iff
$\ObB{C[P]}\sigma = \ObB{C[Q]}\sigma$ for all  contexts~$C$
with $\sigma \gives C[P]\ok$ and $\sigma \gives C[Q]\ok$.
Full abstraction follows by compositionality:

\begin{theorem}[Full abstraction]
  $P =_{\textsc{Den}} Q$ iff $P =_{\textsc{Op}} Q$.
\end{theorem}

\section{Denotational semantics: adding liveness}
\label{sec:liveness}

To round out our study of $\pi$-calculus, we must account for liveness
properties.  Liveness in process algebra appears under diverse guises,
differing in sensitivity to branching behavior and
divergence~\cite{Glabbeek1988}.  Each account of liveness corresponds
to some choice of basic observable: given a process $P$ and a context
$C$, what behavior of $C[P]$ matters?  

The standard observable for the $\pi$-calculus is barbed
bisimilarity~\cite{barbed}, which sits quite far on the branching side
of the linear-branching time spectrum~\cite{Glabbeek1988}.  Here, we
choose a treatment more in the spirit of linear time: an adaptation of
acceptance traces~\cite{Hennessy2002}.  This choice is partly a matter
of taste, but it also allows us to stick with a purely trace-theoretic
semantics, which keeps the domain theory to a minimum.  We do not see
any immediate obstacles to applying our resource-based handling of
names to a branching-time semantics.  Branching sensitivity and
resource-sensitivity seem largely orthogonal, though of course
branches may be pruned when deemed impossible given the owned
resources.

\subsection{Liveness observables}

We say that a process \emph{diverges} if it \emph{can} perform an
infinite sequence of unobservable (\emph{i.e.}, internal) steps
without any intervening interactions with its environment---which is
to say, the process can livelock.  On the other hand, a process that
can make \emph{no} further unobservable steps is blocked (waiting for
interaction from its environment).


The basic observables in our model are:
\begin{itemize}
\item A finite sequence of interactions, after which the process
  diverges or faults;
\item A finite sequence of interactions, after which the process
  is blocked, along with which channels it is blocked on; and
\item An infinite sequence of interactions.
\end{itemize}
\noindent
Notice that we have conflated divergence and faulting: we view both as
erroneous behavior.  In particular, we view any processes that are
capable of immediately diverging or faulting as equivalent, regardless
of their other potential behavior.  This perspective is
reasonable---meaning that it yields a congruence---because such
behavior is effectively uncontrollable.  For example, if $P$ can
immediately diverge, so can $P|Q$ for any $Q$.

Formally, we add a new action $\delta_\Delta$ which records that a
process is blocked attempting
communication along the finite set of \emph{directions} $\Delta$:
\[
\alpha\ ::=\ \cdots \GA \delta_\Delta \qquad
\Delta \subseteq_\textrm{fin} \textsc{Dir} \eqdef 
  \{ c!\ :\ c\in\textsc{Chan} \} \cup \{ c?\ :\ c\in\textsc{Chan} \}
\]
We then define
\[ 
\textsc{LTrace}\ \eqdef\ \textsc{NTAction}^*;\{ \fault, \delta_\Delta\}\ \cup\  
  \textsc{NTAction}^\omega
  \qquad
\textsc{LBeh}\ \eqdef\ \Sigma \rightarrow 2^{\textsc{LTrace}} 
\]
where \textsc{NTAction} (for ``non-terminating action'') refers to all
actions except for $\fault$ or blocking actions $\delta_\Delta$.  Thus
finite liveness traces must end with either a $\delta_\Delta$ action
or a $\fault$ action, whereas neither of these actions can appear in
an infinite trace.

Each liveness trace encompasses a \emph{complete} behavior of the
process: either the process continues interacting indefinitely,
yielding an infinite trace, or diverges, faults or gets stuck after a
finite sequence of interactions.  Therefore, sets of liveness traces
are not prefixed-closed.  

As with the safety traces, we can observe liveness traces from the
operational semantics.  However, we do so using the \emph{greatest}
fixpoint of the following rules:
\begin{display}[$\LObB{P} : \textsc{LBeh}$]{Liveness observation}
\[
\begin{array}{c}
\infer[\tr{LOStep}]
  {P,\sigma \rstep{\alpha }P',\sigma' \\\\
   \alpha \neq \fault \\
   t \in \LOb{P'}{\sigma'}}
  {|\alpha|_\sigma t \in \LOb{P}{\sigma}}
\textrm{gfp}
\qquad
\infer[\tr{LOFault}]
  {P,\sigma \rstep{\fault}}
  {\fault \in \LOb{P}{\sigma}}
\textrm{gfp}
\qquad
\infer[\tr{LOBlocked}]
  {P,\sigma \blocked \Delta}
  {\delta_\Delta \in \LOb{P}{\sigma}}
\textrm{gfp}
\end{array}
\]
\end{display}
where $P, \sigma \blocked \Delta$ means that $P, \sigma$ can only take
communication steps, and $\Delta$ contains precisely the directions of
available communication.  Since the owned resources influence which
communications are possible, they also influence the directions on
which a process is blocked:
\[
\delta_{\{c!\}} \in \LOb{\ov{c}c.0}{[c \mapsto \pub]}
\qquad
\delta_\emptyset \in \LOb{\ov{c}c.0}{[c \mapsto \pri]}
\]
The action $\delta_\emptyset$ reflects a completely deadlocked
process, and is for example the sole trace of the inert process $0$.

Defining the observations via a greatest fixpoint allows for infinite
traces to be observed, but also means that if a process diverges after
a trace $t$, its behavior will contain all traces $tu$, in particular
$t\fault$.  For example, suppose $P,\sigma \rstep{\tau} P,\sigma$.  If
$t$ is any liveness trace whatsoever, we can use the first inference
rule to show, coinductively, that $t \in \LOb{P}{\sigma}$.  We merely
assume that $t \in \LOb{P}{\sigma}$, and derive that $|\tau|_\sigma t
= t \in \LOb{P}{\sigma}$.  Thus, divergence is ``catastrophic'' (as in
failures/divergences~\cite{Brookes1984}).

An important step toward making these observables coherent is the
notion of \emph{refinement}.  In general, saying that $P$ refines $Q$
(or $P$ ``implements'' $Q$) is to say that every behavior of $P$ is a
possible behavior of $Q$.  In other words, $P$ is a more deterministic
version of $Q$.  We define a refinement order on traces:
\[
t \refines t \qquad
t\delta_\Delta \refines t\delta_{\Delta'}\ \textrm{if}\ \Delta' \subseteq \Delta \qquad
tu \refines t\fault
\]
which we lift to sets of traces as: $T \refines U$ iff $\forall t \in
T.\ \exists u\in U.\ t \refines u$.  This notion of refinement, which
closely follows that of acceptance traces~\cite{Hennessy2002}, says
that an implementation must allow at least the external choices that
its specification does.  It also treats faulting as the most
permissive specification: if $Q$ faults, then any $P$ will refine $Q$.
Moreover, any two immediately-faulting processes are equivalent.
Since faulting and divergence are treated identically, the same holds
for divergent processes.  Thus, the simple refinement ordering on
traces has an effect quite similar to the closure conditions imposed
in failures/divergences semantics.

The ordering on trace sets inherits the complete lattice structure of
$2^{\textsc{LTrace}}$, as does the pointwise order on \textsc{LBeh}.
We again exploit this fact when interpreting recursion.

\subsection{Liveness semantics}

To complete the semantic story, we need to interpret blocking actions.
We define
\begin{eqnarray*}
 \ASem{\delta_\Delta}{\sigma} &\eqdef& 
    \begin{cases}
      \top & \exists c.\ (c! \in \Delta \vee c? \in \Delta) \wedge c \notin\dom(\sigma) \\
      \sigma & \ow
    \end{cases} \\
  |\delta_\Delta|_\sigma &\eqdef&
    \delta_{\Delta'} \ \textrm{where}\ 
    \Delta' = \Delta\upharpoonright \{c\ :\ \sigma(c) = \pub \}
\end{eqnarray*}
which shows the interaction between resources and blocking: blocking
on a private resource is possible, but unobservable (\emph{cf.}
projection on $\delta$ in~\cite{Brookes2002}).  For example, we have
\[
\begin{array}{c@{\qquad}c}
\ASem{\delta_{\{c!\}}}{[c \mapsto \pub]} = [c \mapsto \pub] &
|\delta_{\{c!\}}|_{[c \mapsto \pub]} = \delta_{\{c!\}}
\\
\ASem{\delta_{\{c!\}}}{[c \mapsto \pri]} = [c \mapsto \pri] &
|\delta_{\{c!\}}|_{[c \mapsto \pri]} = \delta_\emptyset
\end{array}
\]

The denotational semantics for liveness, $\LSemB{-}$, is largely the
same as that for safety, except for the following clauses:
\begin{eqnarray*}
\LSem{\rec X.P}{\rho}
  &\eqdef&\ \nu B. \LSem{P}{\rho[X \mapsto B]} \\
\LSem{\sum \pi_i. P_i}{\rho} 
  &\eqdef& \left(\biglub \LSem{\pi_i. P_i}{\rho} \right)
    \lub \left(\delta_{\{\dir(\rho\pi_i)\}} \pref \lambda \sigma.\emptyset \right)
\end{eqnarray*}
Recursion is given by a greatest fixpoint, as expected. A summation of
prefixed actions now generates a corresponding blocking set, recording
the external choice (where dir extracts the direction of a prefix).
The blocking action is ``executed'' using the prefixing operator
$\pref$ so that the actual observed action corresponds to the
available resources, as in the example above.

Finally, we use the following definition of interleaving:
\[
\begin{array}{rcll}
t \tparallel u &\eqdef_{\textrm{gfp}}& 
  \alpha \pref (t' \tparallel u) 
    &\IF t = \alpha t',\ \alpha \ \textrm{not blocking} \\&\lub&
  \alpha \pref (t \tparallel u') 
    &\IF u = \alpha u',\ \alpha \ \textrm{not blocking} \\&\lub&
  \delta_{\Delta \cup \Delta'} 
    &\IF t = \delta_\Delta,\ u = \delta_{\Delta'},\ \ov{\Delta} \pitchfork \Delta' \\&\lub&
  t' \tparallel u' &\IF t = \alpha t',\ u = \ov{\alpha} u'
\end{array}
\]
Liveness interleaving is given by a greatest fixpoint.  An infinite
sequence of internal communications (operationally, an infinite
sequence of $\tau$ moves) therefore yields \emph{all} possible traces,
including faulting ones, as it should.  An interleaved trace is
blocked only when both underlying traces are, and only when they do
not block in opposite directions ($\ov{\Delta}$ is $\Delta$ with
directions reversed, and $\pitchfork$ denotes empty intersection).  If
two processes are blocked in opposite directions, then their parallel
composition is in fact \emph{not} blocked, since they are willing to
communicate with each other (\emph{cf} stability~\cite{Brookes1984}).

\subsection{Full abstraction}

The proof of full abstraction is structured similarly to the proof for
the safety semantics.  Congruence proofs must take into account
blocking actions, which is straightforward in all cases except for
parallel composition.  There, we require a lemma:

\begin{lemma}[Blocking congruence]
Suppose $\CI(\sigma_{\textrm{op}}, \sigma_{\textrm{den}}, \sigma_1, \sigma_2)$.  Then
\begin{itemize}
\item If $\delta_{\Delta_i} \in \LOb{P_i}{\sigma_i}$ and $\Delta_1
  \pitchfork \ov{\Delta_2}$ then $|\delta_{\Delta_1 \cup \Delta_2}|_{\sigma_{\textrm{den}}} \in
  \LOb{P_1|P_2}{\sigma_{\textrm{op}}}$.
\item If $\delta_\Delta \in \LOb{P_1|P_2}{\sigma_{\textrm{op}}}$ then
  $\delta_{\Delta_i} \in \LOb{P_i}{\sigma_i}$ for some $\Delta_1$,
  $\Delta_2$ with $\Delta_1 \pitchfork \ov{\Delta_2}$ and
  $|\delta_{\Delta_1 \cup \Delta_2}|_{\sigma_{\textrm{den}}} =
  \delta_\Delta$.
\end{itemize}
\end{lemma}

Defining $=_{\textsc{LDen}}$ and $=_{\textsc{LOp}}$ analogously to the
safety semantics, we again have full abstraction:
\begin{theorem}[Full abstraction]
  $P =_{\textsc{LDen}} Q$ iff $P =_{\textsc{LOp}} Q$.
\end{theorem}

\section{Logic}
\label{sec:logic}

We now sketch a logic for reasoning about the safety semantics of
processes.  The logic proves \emph{refinement} between open
processes---denotationally, trace containment; operationally,
contextual approximation.  The refinements are qualified by assertions
about owned resources, which is what makes the logic interesting.  The
basic judgment of the logic is $\Gamma \gives p \asm P \refines Q$,
which says the traces of $P$ are traces of $Q$, as long as the initial
resources and environment, respectively, satisfy assertions $p$ and $\Gamma$ (defined below).

Resource assertions $p$ are as follows:
\[
p\ ::=\
  \true \GA \false \GA p \wedge q \GA p \vee q \GA p * q \GA
  x\Pub \GA x\Pri \GA x=y \GA x \neq y
\]
and we let $x\Known \eqdef x\Pub \vee x\Pri$.  Satisfaction of
assertions depends on both the environment and resources, as in these
illustrative cases:
\[
\begin{array}{lcl}
\rho, \sigma \models x \Pub &\eqdef& \sigma(\rho(x)) = \pub \\
\rho, \sigma \models p_1 * p_2 &\eqdef& \exists \sigma_1,\sigma_2. \sigma = \sigma_1 \uplus \sigma_2 \textrm{ and } \rho, \sigma_i \models p_i
\end{array}
\]
Resource assertions like $x\Pub$ are
intuitionistic~\cite{Reynolds2002}; without deallocation there is no
reason to use the classical reading, which can assert nonownership.
We are using the standard interpretation of separation logic's $*$ as
disjoint separation to enable \emph{sequential} reasoning about
resource transformers in our logic.  Action interpretations
$\ASemB{\alpha}$ are local with respect to $*$, just as they were for
$\parallel$.

Environment assertions $\Gamma$ constrain process variables:
\[
\begin{array}{c}
\Gamma\ ::=\ \emptyset \GA \Gamma,\ (p \asm X \refines P)
\\
\rho \models (p \asm X \refines P) \ \ \eqdef\ \
\forall \sigma.\ (\rho, \sigma \models p) \implies 
  \rho(X)(\sigma) \subseteq \Sem{P}{\rho}{\sigma}
\end{array}
\]

The definition of entailment is thus:
\[
\Gamma \models p \asm P \refines Q\ \ \eqdef\ \
  \forall \rho, \sigma.\  
    (\rho \models \Gamma\ \wedge\
    \rho, \sigma \models p) \implies 
    \Sem{P}{\rho}\sigma \subseteq \Sem{Q}{\rho}\sigma
\]
By qualifying refinements by resource assertions we can incorporate
Hoare logic-like reasoning.  Take, for example, the rule
\[
\infer
  {\Gamma \gives p * (x\Pub \wedge y\Pub) \asm P \refines Q}
  {\Gamma \gives p * (x\Pub \wedge y\Known) \asm \ov{x}y.P \refines \ov{x}y.Q}
\]
for sending over a public channel.  It is a kind of congruence rule,
but we shift resource assumptions for the subprocesses, corresponding
to the Hoare triple
\[
  \{  p * (x\Pub \wedge y\Known) \}\ \ov{x}y\ \{ p * (x\Pub \wedge y\Pub) \}
\]
The syntactic structure of prefixes (rather than sequential
composition) prevents a clean formulation of the logic using Hoare
triples.  This is why the frame $p$ is included, rather than added via
a separate frame rule; we are using ``large'' rather than ``small''
axioms~\cite{OHearn2001}.  A better treatment is possible if we
semantically interpret prefixing as sequential composition, which
requires a variables-as-resources model~\cite{Parkinson}.

For sending over a private channel, we have an axiom: $\ov{x}y.P$
refines \emph{any} process when $x$ is private, because $\ov{x}y.P$ is
stuck.  The corresponding Hoare triple is $\{ x\Pri \wedge y\Known\}
\ \ov{x}y\ \{\false\}$.


Here is a fragment of the logic, focusing on resource-sensitive rules:
\begin{display}[$\Gamma \gives p \asm P \refines Q$]
{A selection of logical rules for safety behavior}
\[
\infer
  {\Gamma \gives p * (x\Pub \wedge y\Pub) \asm P \refines Q}
  {\Gamma \gives p * (x\Pub \wedge y\Known) \asm \ov{x}y.P \refines \ov{x}y.Q}
\quad
\infer
  { }
  {\Gamma\gives x\Pri \wedge y\Known \asm \ov{x}y.P \refines Q}
\]
\[
\infer
  {\Gamma \gives (p * x\Pub) \wedge y\Pub \asm P \refines Q \\
   y \notin \fv(p, \Gamma)}
  {\Gamma \gives p * x\Pub \asm x(y).P \refines x(y).Q}
\quad
\infer
  { }
  {\Gamma \gives x\Pri \asm x(y).P \refines Q}
\]
\[
\infer
  {\Gamma \gives p * x\Pri \asm P \refines Q \\
   x \notin \fv(p, \Gamma)}
  {\Gamma \gives p \asm \new x.P \refines \new x.Q}
\qquad
\infer
  {\Gamma \gives \widehat{p} \asm P_i \refines Q_i}
  {\Gamma \gives p \asm P_1|P_2 \refines Q_1|Q_2}
\]
\[
\infer
  {p \asm X \refines P \in \Gamma }
  {\Gamma \gives p \asm X \refines P}
\qquad
\infer
  {\Gamma, p \asm X \refines Q \gives p \asm P \refines Q}
  {\Gamma \gives p \asm \rec X. P \refines Q}
\qquad
\infer
  {p \models p' \\
   \Gamma \gives p' \asm P \refines Q}
  {\Gamma \gives p \asm P \refines Q}
\]
\end{display}
The congruence rule for parallel composition performs public-lifting
$\widehat{p}$ on resource assertions (by replacing $\pri$ by $\pub$ in
the assertion).

Fixpoint induction is resource-qualified as well.  We reason about the
body $P$ of a recursive definition $\rec X.P$ using a hypothetical
bound on $X$ as the induction hypothesis.  That hypothesis, however,
is only applicable under the \emph{same} resource assumptions $p$ that
were present when it was introduced---making $p$ the loop invariant.

In addition to these resource-sensitive rules, we have the usual laws
of process algebra, including the expansion law.  Combining those laws
with the ones we have shown, we can derive an \emph{interference-free}
expansion law, as in this simplified version:
$
\Gamma \gives x\Pri \wedge y\Known \asm
  \ov{x}y.P | x(z).Q \equiv
  P | Q\{y/z\}
$.


\section{Discussion}

\subsection{Future work: richer resources}
\label{sec:resources}

Our resource model captures exactly the guarantees provided by the
$\pi$-calculus: until a channel is exposed, it is unavailable to the
environment; afterwards, all bets are off.  This property is reflected
in the fact that $\Sigma$ is not a separation algebra, since $c\ \pub
\parallel c\ \pub$ can result in $c\ \pub$ or $c\ \pri$.  No amount of
public ownership adds up definitively to private ownership.

Rather than using resources to model the guarantees of a language, we
can instead use them to enforce guarantees we intend of programs, putting
ownership ``in the eye of the asserter''~\cite{O'Hearn2007}.  We can
then recover privacy just as Boyland showed~\cite{Boyland2003} how to
recover write permissions from read permissions: via a
fractional model of ownership: $ \Sigma_{\textsc{Frac}} \eqdef
\textsc{Chan} \ra [0, 1] $.  Unlike traditional fractional
permissions, owning a proper fraction of a channel does not limit what
can be done with the channel---instead, it means that the environment
is \emph{also} allowed to communicate on the channel.  The fractional
model yields a separation algebra, using (bounded) summation for
resource addition.  An easy extension is distinguishing send
and receive permissions, so that interference can be ruled out in a
direction-specific way.

One can also imagine encoding a session-type discipline~\cite{Honda1998} as a
kind of resource: $\Sigma_\textsc{Sess} \eqdef \textsc{Chan}
\rightharpoonup \textsc{Session}$ where
\[
   s \in \textsc{Session}\ ::=\ 
     \ell.s \oplus \ell.s \GA \ell.s\ \&\ \ell.s \GA !.s \GA ?.s \GA \textsf{end} 
\]
Separation of session resources corresponds to matching up dual
sessions, and actions work by consuming the appropriate part of the
session.  Ultimately, such resource models could yield rely-guarantee
reasoning for the $\pi$-calculus, borrowing ideas from
deny-guarantee~\cite{Dodds2009}.  A challenge for using these models
is managing the ownership protocol in a logic: how are resources
consistently attached to channels, and how are resources split when
reasoning about parallel composition?  We are far from a complete
story, but believe our semantics and logic can serve as a foundation
for work in this direction.

\subsection{Related work}
\label{sec:related}

Hoare and O'Hearn's work~\cite{Hoare2008} introduced the idea of
connecting the model theory of separation logic with the
$\pi$-calculus, and provided the impetus for the work presented here.
Their work stopped short of the full $\pi$-calculus, modelling only
point-to-point communication and only safety properties.  Our liveness
semantics, full abstraction results, and refinement calculus fill out
the rest of the story, and they all rely on our new resource model.
In addition, our semantics has clearer connections to both Brookes's
action trace model~\cite{Brookes2002} and abstract separation
logic~\cite{Calcagno2007}.

Previous fully abstract models of the $\pi$-calculus are based on
functor categories~\cite{Stark2002,Hennessy2002,Fiore2002}, faithfully
capturing the traditional role of scope for privacy in the
$\pi$-calculus.  Those models exploit general, abstract accounts of
recursion, nondeterminism, names and scoping in a category-theoretic
setting.  We have similarly sought connections with a general
framework, but have chosen resources, separation and locality as our
foundation.  

An immediate question is: why do we get away with so much less
mathematical scaffolding?  This question is particularly pertinent in
the comparison with Hennessy's work~\cite{Hennessy2002}, which uses a
very similar notion of observation.  Hennessy's full abstraction
result is proved by extracting, from his functor-categorical
semantics, a set of acceptance traces, and showing that this
extraction is injective and order preserving.  The force of this
``internal full abstraction'' is that the functor-categorical meaning
of processes is completely determined by the corresponding acceptance
traces.  But note, these traces are \emph{not} given directly via a
compositional semantics: they are extracted only after the
compositional, functor-categorical semantics has been applied.  What
we have shown, in a sense, is that something like acceptance traces
for a process can be calculated directly, and compositionally, from
process syntax.

Beyond providing a new perspective on the $\pi$-calculus,
we believe the resource-oriented approach will yield new reasoning
techniques, as argued above.  We have also emphasized concreteness,
giving an elementary model theory based on sets of traces.

Finally, it is worth noting that substructural type systems have been
used to derive strong properties (like confluence) in the
$\pi$-calculus~\cite{Kobayashi1999}, just as we derived
interference-free expansion.  Here, we have used a resource theory to
explain the $\pi$-calculus as it is, rather than to enforce additional
discipline.  But the ideas of~\secref{resources} take us very much
into the territory of discipline enforcement.  More work is needed to
see what that territory looks like for the resource-based approach.

\paragraph{Acknowledgements}  We are grateful to Paul Stansifer and 
Tony Garnock-Jones for feedback on drafts of this paper, and to the
anonymous reviewers who provided guidance on presentation.

\bibliography{pi-semantics}{} \bibliographystyle{entcs}

\begin{thebibliography}{10}
\expandafter\ifx\csname url\endcsname\relax
  \def\url#1{\texttt{#1}}\fi
\expandafter\ifx\csname urlprefix\endcsname\relax\def\urlprefix{URL }\fi
\newcommand{\enquote}[1]{``#1''}

\bibitem{Boyland2003}
Boyland, J., \emph{{Checking Interference with Fractional Permissions}}, in:
  \emph{SAS}, 2003.

\bibitem{Brookes2002}
Brookes, S., \emph{{Traces, Pomsets, Fairness and Full Abstraction for
  Communicating Processes}}, in: \emph{CONCUR}, 2002, pp. 45 --71.

\bibitem{Brookes2007}
Brookes, S., \emph{{A semantics for concurrent separation logic}}, TCS
  \textbf{375} (2007), pp.~227--270.

\bibitem{Brookes1984}
Brookes, S.~D. and A.~W. Roscoe, \emph{{An Improved Failures Model for
  Communicating Processes}}, in: \emph{Seminar on Concurrency}, 1984.

\bibitem{Calcagno2007}
Calcagno, C., P.~W. O'Hearn and H.~Yang, \emph{{Local Action and Abstract
  Separation Logic}}, in: \emph{LICS}, 2007.

\bibitem{Dodds2009}
Dodds, M., X.~Feng, M.~Parkinson and V.~Vafeiadis, \emph{{Deny-guarantee
  reasoning}}, in: \emph{ESOP}, 736 (2009), pp. 363--377.

\bibitem{Fiore2002}
Fiore, M., E.~Moggi and D.~Sangiorgi, \emph{{A fully-abstract model for the
  pi-calculus}}, in: \emph{LICS}, December (1996).

\bibitem{Hennessy2002}
Hennessy, M., \emph{{A fully abstract denotational semantics for the
  pi-calculus}}, TCS \textbf{278} (2002), pp.~53--89.

\bibitem{Hoare2008}
Hoare, T. and P.~O'Hearn, \emph{{Separation Logic Semantics for Communicating
  Processes}}, Electronic Notes in Theoretical Computer Science (ENTCS)
  (2008).

\bibitem{Honda1998}
Honda, K., V.~T. Vasconcelos and M.~Kubo, \emph{{Language Primitives and Type
  Discipline for Structured Communication-Based Programming}}, in: \emph{ESOP},
  1998, pp. 122--138.

\bibitem{Kobayashi1999}
Kobayashi, N., B.~Pierce and D.~Turner, \emph{{Linearity and the pi-calculus}},
  ACM Transactions on Programming Languages and Systems (TOPLAS) \textbf{21}
  (1999), pp.~914--947.

\bibitem{Milner1992}
Milner, R., J.~Parrow and D.~Walker, \emph{{A calculus of mobile processes,
  parts I and II}}, Information and computation \textbf{100} (1992).

\bibitem{barbed}
Milner, R. and D.~Sangiorgi, \emph{Barbed bisimulation}, in: \emph{Automata,
  Languages and Programming},  Lecture Notes in Computer Science  \textbf{623},
  1992 pp. 685--695.

\bibitem{O'Hearn2007}
O'Hearn, P., \emph{{Resources, concurrency, and local reasoning}}, TCS
  \textbf{375} (2007), pp.~271--307.

\bibitem{OHearn2001}
O'Hearn, P., J.~Reynolds and H.~Yang, \emph{{Local Reasoning about Programs
  that Alter Data Structures}}, in: \emph{Computer Science Logic}, 2001.

\bibitem{Parkinson}
Parkinson, M., R.~Bornat and C.~Calcagno, \emph{{Variables as Resource in Hoare
  Logics}}, in: \emph{LICS} (2006).

\bibitem{Reynolds2002}
Reynolds, J., \emph{{Separation logic: a logic for shared mutable data
  structures}}, in: \emph{LICS}, 2002.

\bibitem{Roscoe1993}
Roscoe, A.~W. and G.~Barrett, \emph{{Unbounded Non-determinism in CSP}}, in:
  \emph{MFPS}, 1989.

\bibitem{Sangiorgi2001}
Sangiorgi, D. and D.~Walker, \enquote{{The pi-calculus: a Theory of Mobile
  Processes},} Cambridge University Press, 2001.

\bibitem{Stark2002}
Stark, I., \emph{{A fully abstract domain model for the pi-calculus}}, in:
  \emph{LICS} (1996), pp. 36--42.

\bibitem{Glabbeek1988}
{Van Glabbeek}, R., \emph{{The linear time-branching time spectrum}}, CONCUR'90
  Theories of Concurrency: Unification and Extension  (1990), pp.~278--297.

\end{thebibliography}

\end{document}